\documentclass[11pt,a4paper]{article}
\pdfoutput=1
\usepackage{jcappub}
\usepackage[T1]{fontenc} 
\usepackage[utf8]{inputenc}
\usepackage{graphicx}
\usepackage{lineno}  

\title{\boldmath Measurement of the cosmic ray muon flux seasonal variation with the OPERA detector}

\author[a]{N. Agafonova,}
\author[b]{A. Alexandrov,}
\author[c]{A. Anokhina,}
\author[d]{S. Aoki,}
\author[e]{A. Ariga,}
\author[e,f]{T. Ariga,}
\author[g]{A. Bertolin,}
\author[h]{C. Bozza,}
\author[g,i]{R. Brugnera,}
\author[b,j,1]{A. Buonaura,\note[$\star$]{Corresponding authors: nicoletta.mauri@bo.infn.it, alessandro.paoloni@lnf.infn.it.}\note{Now at Physik-Institut, Universit{\"a}t Z{\"u}rich, Z{\"u}rich, Switzerland.}}
% ...

\affiliation[a]{INR - Institute for Nuclear Research of the Russian Academy of Sciences, RUS-117312 Moscow, Russia}
\affiliation[b]{INFN Sezione di Napoli, I-80126 Napoli, Italy}
\affiliation[c]{SINP MSU - Skobeltsyn Institute of Nuclear Physics, Lomonosov Moscow State University, RUS-119991 Moscow, Russia}
\affiliation[d]{Kobe University, J-657-8501 Kobe, Japan}
\affiliation[e]{Albert Einstein Center for Fundamental Physics, Laboratory for High Energy Physics (LHEP), University of Bern, CH-3012 Bern, Switzerland}
\affiliation[f]{Faculty of Arts and Science, Kyushu University, J-819-0395 Fukuoka, Japan}
\affiliation[g]{INFN Sezione di Padova, I-35131 Padova, Italy}
\affiliation[h]{Dipartimento di Fisica dell'Universit\`a di Salerno and ``Gruppo Collegato''  INFN, I-84084 Fisciano (Salerno), Italy}
\affiliation[i]{Dipartimento di Fisica e Astronomia dell'Universit\`a di Padova, I-35131 Padova, Italy}
\affiliation[j]{Dipartimento di Fisica dell'Universit\`a Federico II di Napoli, I-80126 Napoli, Italy}
% ...

\abstract{The OPERA experiment discovered 
muon neutrino into tau
neutrino oscillations in appearance mode, detecting tau leptons by means of
nuclear emulsion films. The apparatus was also
endowed with electronic detectors with 
tracking capability, such
as scintillator strips and resistive plate chambers.  Because of its
location, in the underground Gran Sasso laboratory, under 3800 m.w.e.,
the OPERA detector  
has also been used as an observatory for TeV muons produced by
cosmic rays in the atmosphere.  In this paper the measurement of the
single muon flux modulation and of its correlation with the seasonal
variation of the atmospheric temperature are reported.}

\begin{document}
\maketitle
\flushbottom

\section{Introduction}
\label{sec:Intro}
Muons observed in underground laboratories are produced mainly in
decays of pions and kaons generated by the interaction of primary
cosmic rays with the upper atmosphere.  Since muons loose energy
crossing the rock overburden, only high energy muons can be detected, 
with an energy threshold depending on depth, 
usually expressed in metre water equivalent (m.w.e.).  
It is known that the flux of atmospheric muons detected deep underground shows 
seasonal time variations correlated with the temperature of the stratosphere, 
where the primary cosmic rays interact~\cite{bib:season}. 
This effect has been reported by other experiments at the underground Gran Sasso Laboratory (LNGS) 
(MACRO~\cite{bib:macro1, bib:macro1bis, bib:macro2},
LVD~\cite{bib:lvd}, Borexino~\cite{bib:borex, bib:borex2} and GERDA~\cite{bib:gerda}) 
and elsewhere (AMANDA~\cite{bib:amanda}, IceCube~\cite{bib:icecube}, MINOS~\cite{bib:minosf, bib:minosn}, 
Double Chooz~\cite{bib:db} and Daya Bay~\cite{bib:dayabay}).
An increase in temperature of the stratosphere, indeed, 
causes a decrease of 
the air density, thus reducing the chance of $\pi, K$ mesons to interact, and 
resulting in a larger fraction decaying into muons. 
So, the atmospheric muon rate changes during the year 
increasing in summer and decreasing in winter. 
The variation can be modeled as a sinusoidal function, which is only a first 
order approximation since the average temperature is not precisely constant 
over the years and short term effects occur, leading to local maxima and 
minima, like the ``sudden stratospheric warming'' events~\cite{bib:ssw}. 

The OPERA experiment~\cite{bib:detector} discovered 
$\nu_\mu$ into $\nu_\tau$ oscillations in appearance mode
using the CERN Neutrino to Gran Sasso (CNGS) beam~\cite{bib:prl2015, bib:prl2018}.  
The experiment was
located 
in the Hall C of 
LNGS laboratory, at 3800 m.w.e. depth.
To identify $\nu_\tau$ Charged Current interactions, the Emulsion Cloud Chamber 
technique was used, with 1 mm lead sheets alternated with 
nuclear emulsion 
films, for a total target mass of about 1.2 kt. 
Lead and emulsions were organised into units called ``bricks''. 
The detector was divided into two identical Super-Modules (SM), each made of a target section, 
composed by 31 brick walls interleaved with target tracker layers to locate bricks with neutrino interactions, 
and of a muon
spectrometer to optimise the muon identification probability and to
measure momentum and charge. 
In each SM, the Target Tracker (TT)~\cite{bib:TTref} 
consisted of 31 pairs of orthogonal planes made of 
2.6~cm wide 
scintillator strips, 
read-out by means of 
WLS fibers. 
The fibers from the 6.7~m long strips were collected into four
groups of 64 and coupled to 64-channel Hamamatsu H7546
photomultipliers.
The muon 
spectrometers were made by an iron-core dipole magnet with
drift tubes used as precision trackers and 22 layers of resistive
plate chambers (RPC)~\cite{bib:RPC} inside the magnetised iron. 
The 2D read-out was performed by
means of 2.6~cm pitch and 8~m long vertical strips, which measured the
coordinate in the bending plane, and 3.5~cm pitch and 8.7~m long
horizontal strips, measuring the orthogonal coordinate.  

The analysis presented here is based on TT and RPC data recorded
during about five years from January 2008 to March 2013. 
In the TT, 
the trigger condition required either hits in the horizontal and 
vertical views of at least two 
planes or 
at least 10 
hits in a single plane 
with a signal greater than 
30 photo-electrons for 2008 and 2009 runs;  
from 2010 up to 2013 
the latter requirements were lowered to 
4 hits and 10 
photo-electrons, 
respectively. 
Data from the RPCs of each
spectrometer were acquired 
in presence of at least 3 planes fired in a
time window of 200~ns.  
Events were recorded in presence of at least 5
TT and/or RPC hits in each view, horizontal and vertical, within a time window of 500~ns.

The TT systems were operative during most of the time in
the considered years, while the RPCs had a lower run-time, being
operative only during CNGS runs 
and switched off during the CNGS
winter shutdowns. 
More details about the electronic 
detectors used for the cosmic ray muon analysis can be found in
\cite{bib:detector}.

\section{Cosmic ray muon flux measurement and its modulation}
\label{sec:Cray}
Cosmic ray induced events in the OPERA detector were selected, through 
their absolute time, 
outside of the CNGS spill window.  
Once the event was tagged as ``off-beam'' it was classified as cosmic and 
processed in a dedicated way.  
The 
standard reconstruction package ``OpRec''~\cite{oprec} 
was complemented with a set of algorithms developed for the different
cosmic and beam event topologies. 
The reconstruction was effective at identifying single and multiple muon tracks
(muon bundles). 
For this analysis, a total of about 4 million single muon events have been 
selected requiring a single track reconstructed in both views. 

Different atmospheric muon rates have been measured in periods with and without RPC acquisition. 
The scale factor between the two rates has been evaluated directly from data over the full data taking period, 
extracting the two constant terms 
$I^0_{TT+RPC} = (3359 \pm 5)~\mu$/day and $I^0_{TT-only} = (1960 \pm 5)~\mu$/day from a maximum likelihood fit on the two data sets.

\begin{figure}[]
\centerline{\includegraphics[width=\linewidth]{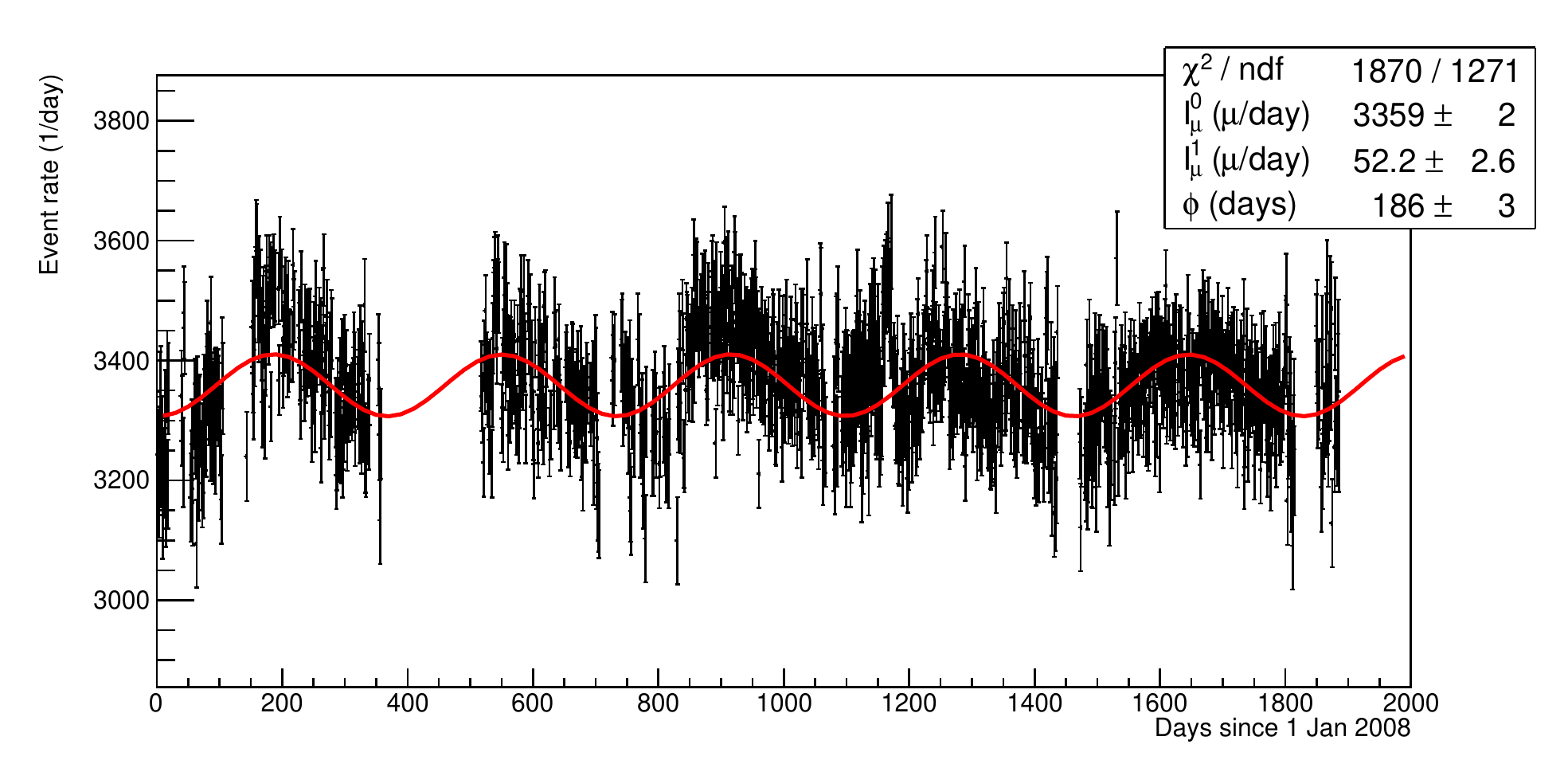}}
\centerline{\includegraphics[width=\linewidth]{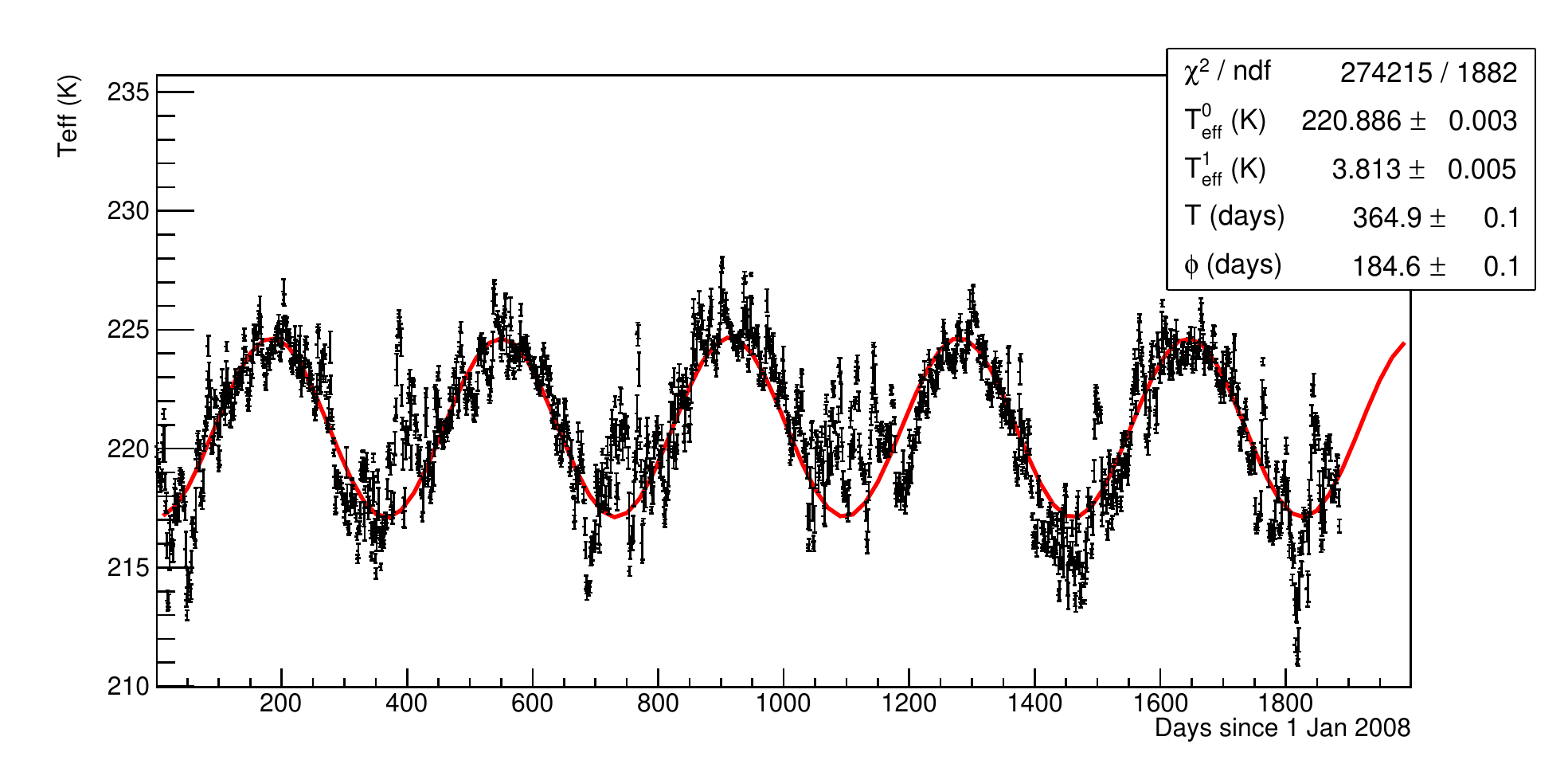}}
\caption[]{Single muon rate measured by the OPERA detector (top) and
effective atmospheric temperature (bottom) from January 2008 to March 2013.
Fit results are superimposed to the data sets. 
Symbols in the legends are defined in the text and in Eqs.~\ref{murate} and~\ref{trate}.
}
\label{fig:murate}
\end{figure}

In Fig.~\ref{fig:murate} (top panel) the flux of atmospheric single muons measured 
from 1 January 2008 (day 1 in the plot) to 
March 2013 is shown. 
After data quality cuts, 
our data set is composed of 1274 live days, out of which 919 days with TT+RPC. 
The longer downtime period corresponds to the first 5 months of 2009, when the acquisition was 
stopped due to a DAQ upgrade, at first, and then to the earthquake in L'Aquila. 
Other shorter downtime periods are present in winter due to maintenance operations. 
Data with TT-only acquisition have been rescaled to the TT+RPC average rate. 
The flux has been fitted to 
\begin{equation}
I_{\mu}(t) = I^0_\mu + I^1_\mu \cos \frac{2 \pi} {T} (t-\phi) 
\label{murate}
\end{equation}
The presence of a sinusoidal 
component with a period $T=(359 \pm 2)$ days is observed, with an
amplitude amounting to $\alpha= I_\mu^1/I_\mu^0 = (1.55 \pm 0.08)$\% of the average flux, 
and a phase $\phi=(197 \pm 5)$~days, with a $\chi^2/\textrm{dof} = 1.46$. 
In accordance with other LNGS experiments and 
with the fit result on the temperature reported in Sec.~\ref{sec:Temperature}, 
we take as the best estimate of the phase the result obtained fixing the period to one year of 365 days. 
The maximum is then observed 
at 
day 
$\phi = (186 \pm 3)$, corresponding to July 5, 
with a $\chi^2$/dof value of 1.47. The result is also shown in Fig.~~\ref{fig:murate} (top panel).

\begin{figure}
\includegraphics[width=0.5\linewidth]{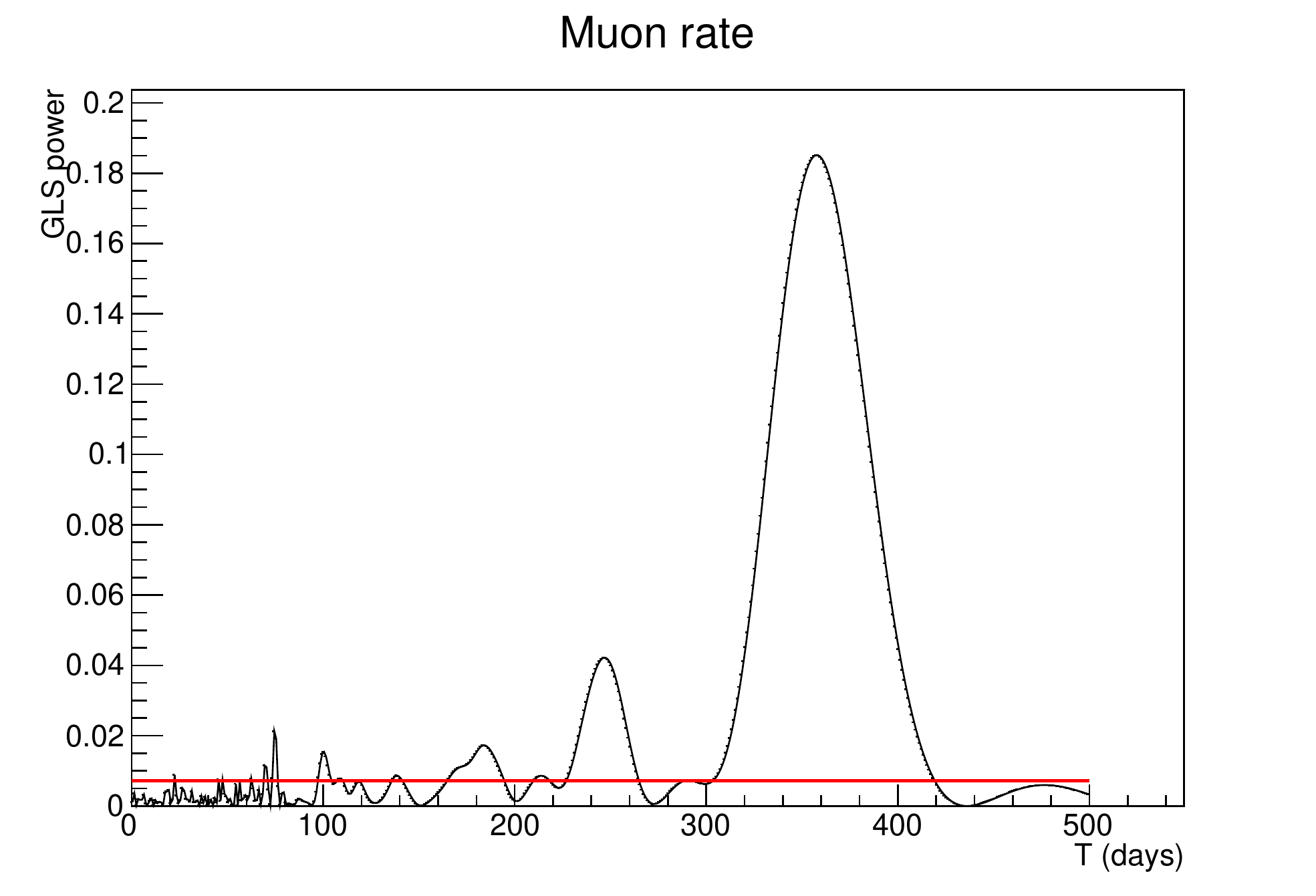}
\includegraphics[width=0.5\linewidth]{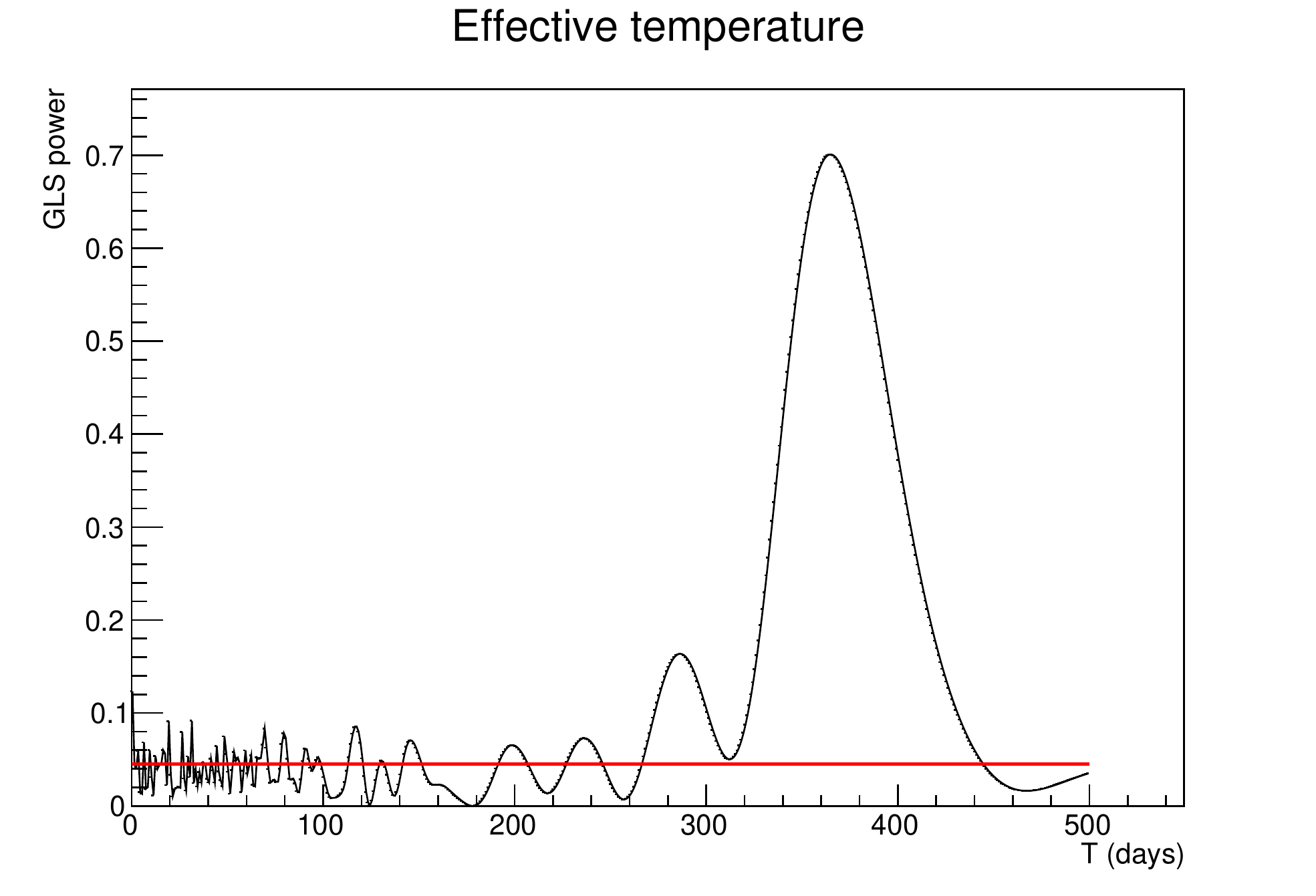}
\caption[]{Generalised Lomb-Scargle periodograms for the measured muon
  rate (left) and the effective atmospheric temperature (right).  
The 99\% significance level is also drawn as a reference in both periodograms. }
\label{fig:LS}
\end{figure}

The Lomb-Scargle 
periodogram \cite{bib:LS1,bib:LS2} is a common tool to analyze 
unevenly spaced data 
to detect a periodic variation 
independently of the modulation phase. 
For the analysis presented here, we have 
exploited the generalised Lomb-Scargle 
periodogram, proposed in \cite{bib:LSgen}, which takes into account the 
non-zero average value of the event rate.  
The periodogram obtained for the single muon event rate is shown in Fig.~\ref{fig:LS} (left panel). 
To assess the significance of the periodogram peaks, 10$^5$ toy
experiments with a constant rate of 3359~$\mu$/day in the 
detector have been simulated, and the corresponding periodograms
reconstructed.  In Fig.~\ref{fig:LS} the 99\% significance level,
defined as the value for which 99\% of the toy experiments result in a
lower spectral power, is also shown.

The most significant peak is around one year ($P_{max}$ at $T \sim 365$ days), but 
other less significant peaks are also present, as a consequence of the fact
that Eq.~\ref{murate} is 
an approximation. 
A simulated experiment has been performed extracting 
daily rates according to the result of our fit and comparing the periodograms obtained 
with and without data in the days of detector downtime. 
It shows that the amplitude of the peaks around 200 days 
increases with the detector downtime. 

We checked possible systematic effects on the phase $\phi$ coming from the data taking stability along the years. 
Possible variations in DAQ efficiency are likely to coincide with calendar years by the fact that RPC were turned on and off on a yearly basis and DAQ maintenance and interventions were done during the CNGS winter shutdowns. 
By applying scale factors on a yearly basis,
a constant average rate is achieved by definition. 
The normalisation, different at the few per mille level, was applied 
to TT-only and TT+RPC data, which were then rescaled one to the other. 
The rate has been fitted to Eq.~\ref{murate} 
fixing the period to 365 days, the modulation maximum results on day~$\phi = (177 \pm 3)$ with a $\chi^2/\textrm{dof} = 1.64$. 
Comparing these results with those obtained with a constant scale factor, 
we evaluate a systematic error on the phase as the semi-difference between 
the two $\phi$ values 
extracted 
with period fixed to one year, 
i.e. $\delta \phi_{\textrm{sys}} = 5$~days. 
Our best estimate of the muon rate maximum, obtained at fixed period $T = 365$~days, is found on day 
$\phi = 186 \pm 3_{\textrm{stat}} \pm 5_{\textrm{sys}}$, i.e. July 5. 

\section{Atmospheric temperature modulation}
\label{sec:Temperature}
To measure the atmospheric temperature modulation, we have used data
from the European Center for Medium-range Weather Forecasts 
(ECMWF)~\cite{bib:ECMWF}.
The center provides temperature values at different altitudes above given
locations, obtained by means of interpolations based on measurements 
of various kinds around 
the planet and on a global atmospheric model. 
The coordinates are those used in \cite{bib:borex}: 
13.5333$^\circ$~E, 42.4275$^\circ$~N.  
The atmospheric temperature is provided at 37 discrete pressure levels 
ranging from 1 to 1000 hPa four times in each day (0.00~h, 6.00~h, 12.00~h 
and 18.00~h).  
Averaging these temperature values using weights accounting for the production
of pions and kaons at different altitudes (see Appendix~\ref{sec:Appendix}), 
the effective atmospheric temperature $T_{eff}$ has been calculated four times 
a day.  
The four measurements are then averaged and the variance used as an estimate 
of the uncertainty on the mean value.

The weights used in this analysis have been computed as described in~\cite{bib:Grashorn} and in 
previous experimental papers~(\cite{bib:macro1}-\cite{bib:dayabay}). They depend on the inclusive meson
production in the forward region, on the attenuation lengths of 
cosmic ray primaries,  
pions and kaons, as well as on the average value 
$\langle E_{thr} \cos \theta \rangle$, where $E_{thr}$ is the minimum energy required
for a muon to reach the considered underground site and $\theta$ is
the angle between the muon and the vertical direction. 
From a full Monte Carlo simulation taking into account the rock map above Hall C of LNGS~\cite{chargeratio}, it results that 
$E_{thr} = 1.4$~TeV and $\langle E_{thr} \cos \theta \rangle = (1.1 \pm 0.2)$~TeV.   
A previous analysis~\cite{bib:borex} quoted a higher value, $E_{thr} = 1.8$~TeV,
extracted from numerical methods 
assuming a flat overburden~\cite{bib:Grashorn}. 
All other parameters in the weight functions are site independent.
More details about the effective atmospheric temperature calculation are 
given in Appendix~\ref{sec:Appendix}.

The $T_{eff}$ values are shown in Fig.~\ref{fig:murate}~(bottom panel) in the 
time period of the 
data taking, from January 2008 to March 2013.
The temperature has been fitted to a sinusoidal function similar to Eq.~\ref{murate}: 
\begin{equation}
T_{eff}(t) = T^0_{eff} + T^1_{eff} \cos \frac{2 \pi} {T} (t-\phi) 
\label{trate}
\end{equation}
The fit results are also shown in Fig.~\ref{fig:murate}. 
The average effective temperature is 220.9~K, and a modulation is 
observed with an amplitude $\alpha = T_{eff}^1/T_{eff}^0 = (1.726 \pm 0.002)\%$ of $T^0_{eff}$. 
The period $T = (364.9 \pm 0.1)$~days and phase $\phi=(184.6 \pm 0.1)$~days are similar to those observed for the
single muon rate. 
A more refined study about the time correlation between temperature and
muon rate is 
presented 
in the next Section.
In Fig.~\ref{fig:LS} (right panel) the generalised Lomb-Scargle periodogram is displayed also for 
$T_{eff}$. As for the muon rate, the most significant peak is around 365 days 
and other less significant peaks are present. 

\section{Cosmic ray flux and effective atmospheric temperature correlation}
\label{sec:correlation}
The possible presence of a time shift $\tau$ between the modulated components of
the cosmic ray muon rate 
and of the effective atmospheric temperature has been 
investigated using the cross correlation function defined as:

\begin{equation}
R(\tau)~ =~ \int_0^{\Delta t} \frac{I_\mu (t)-I^0_\mu}{I^1_\mu}~\frac{T_{eff}(t-\tau)-T^0_{eff}}{T^1_{eff}}~\frac{dt}{\Delta t} ~\simeq~ \frac{1}{N_{d}}~\Sigma_i~\frac{I_\mu (t_i)-I^0_\mu}{I^1_\mu}~\frac{T_{eff}(t_i-\tau)-T^0_{eff}}{T^1_{eff}} 
\label{eqtimecor}
\end{equation}
$I^0_\mu$ and $T^0_{eff}$ are the average values as obtained in the fits of 
Sec.~\ref{sec:Cray}, for the single muon cosmic events, and 
of Sec.~\ref{sec:Temperature}, for the effective atmospheric temperature, while
$I^1_\mu$ and $T^1_{eff}$ are the corresponding amplitudes of the 
modulated components.
The sum runs over the $N_{d}$ days with both measurements.  

\begin{figure}[]
\centerline{\includegraphics[width=0.7\linewidth]{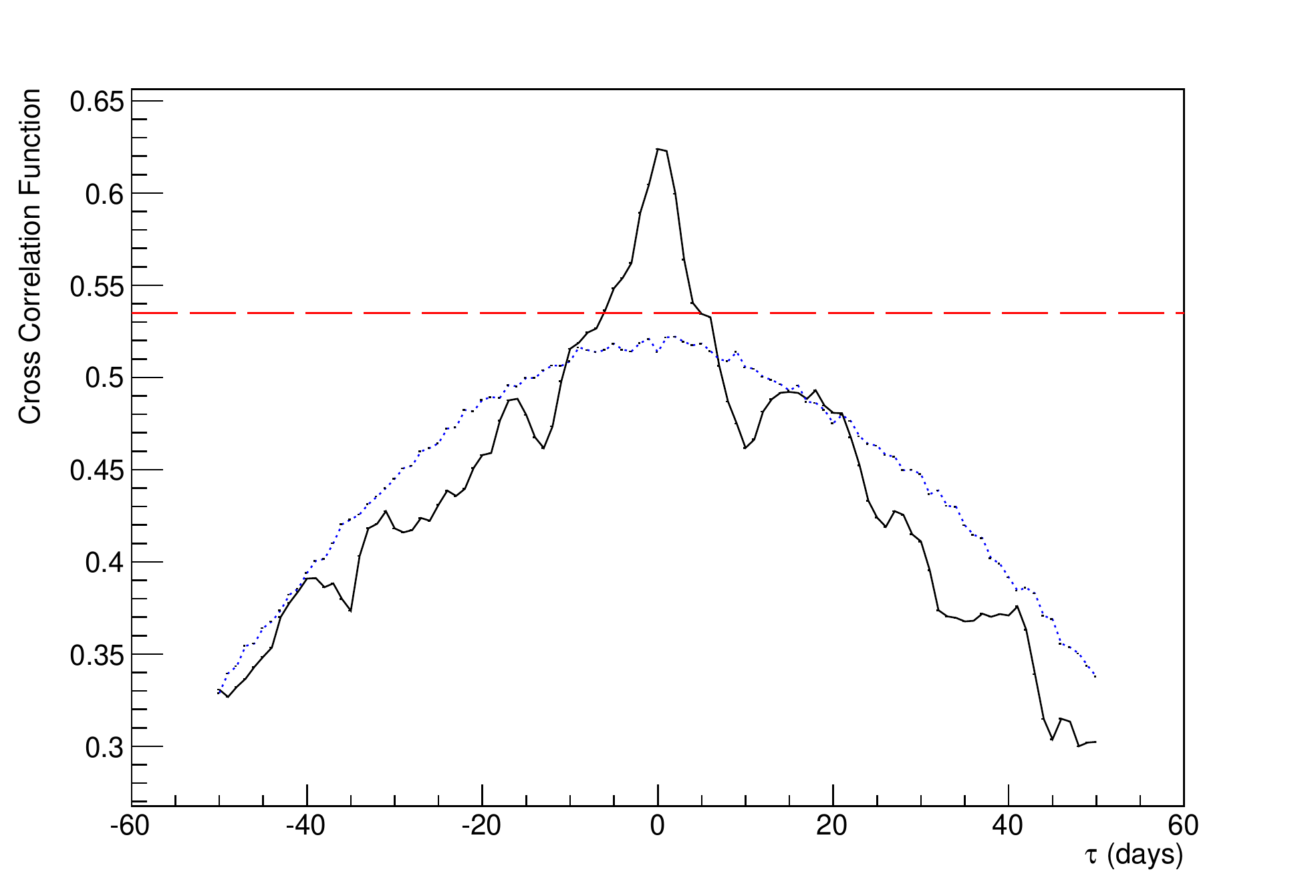}}
\caption[]{Cross correlation function (continuous black) between the measured daily muon rate
  and the effective atmospheric temperature. In dotted blue the result of a toy 
  Monte Carlo simulation described in the text 
  is reported, where the muon rate and the effective 
  temperature have been extracted according to the fit results, but with equal
  time period and phase. In dashed red the 99\% significance level is also
  shown (see text for the definition).}
\label{fig:correphase}
\end{figure}

The correlation function is shown in Fig.~\ref{fig:correphase},
together with the 99\% C.L., in dashed red, evaluated by producing with Monte Carlo
techniques 10$^5$ toy experiments, each one consisting of two time
series, one for the temperature and the other for the muon rate.  The
99\% C.L. value is defined as that for which 99\% of the toy
experiments have a correlation value for $\tau=0$ day lower than it.
Both temperatures and rates have been generated according to the results
of the fits reported in the previous sections, fixing the time period 
to 365 days and the phase to 185 days both for the simulated cosmic ray flux
and effective atmospheric temperature.
In the same figure, in dotted blue, the cross correlation function is reported, as
a reference, for one of the toy experiments.
In real data 
a peak with maximum at $\tau = 0$ day 
is observed 
above the expected 
contribution of the modulated components. 
This peak is due to correlated short term deviations (few
days scale) from the fitted functions in the atmospheric muon rate and the effective temperature. 
Both the sinusoidal components and the short term variations of 
the two 
time series are synchronous. 

\begin{figure}
\centering
\includegraphics[width=0.8\linewidth]{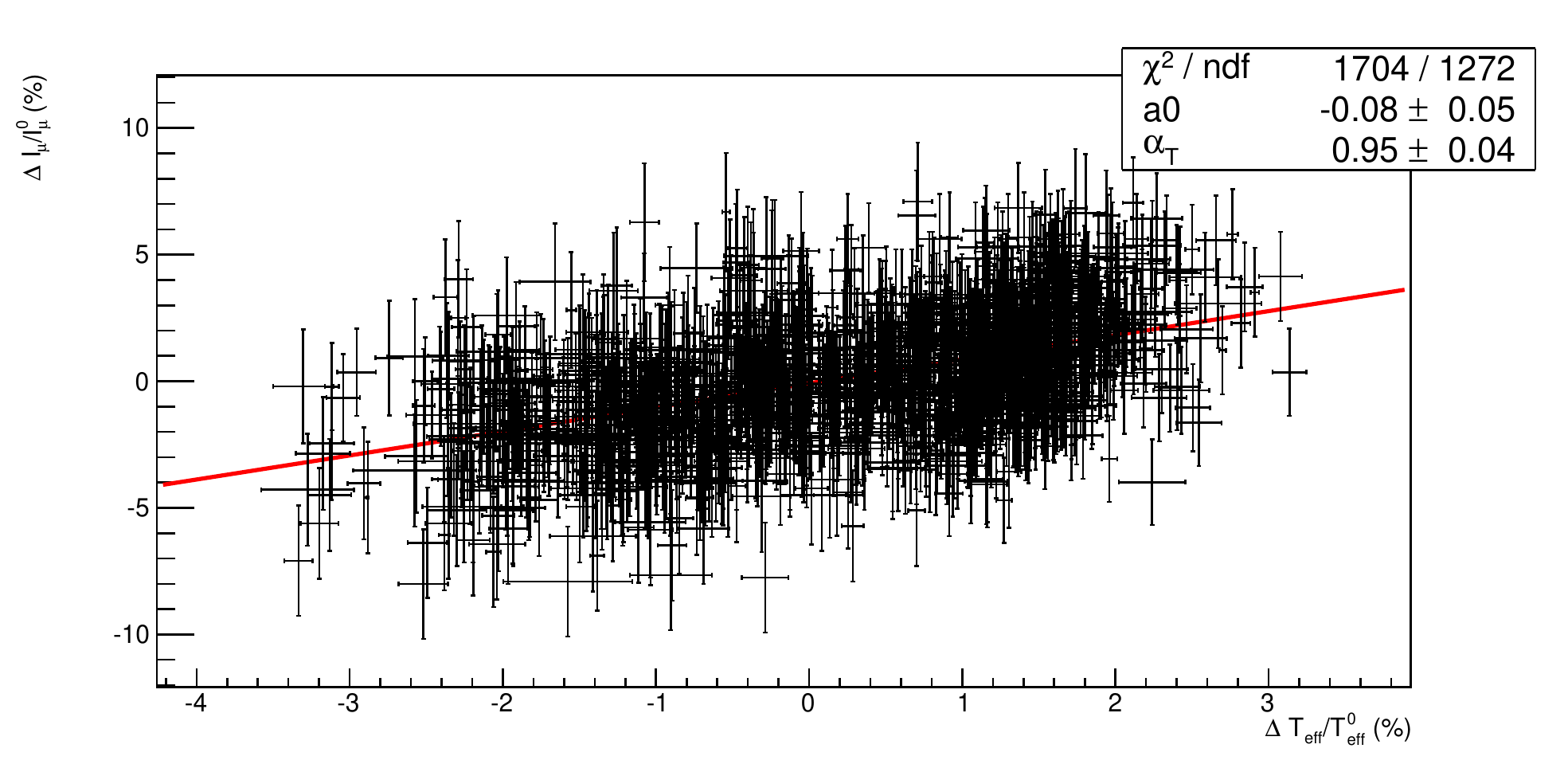}
\caption[]{Correlation between the muon rate and the effective
  temperature relative variations. Symbols in the legend are defined in the text. }
\label{fig:alphaT}
\end{figure}

In Fig.~\ref{fig:alphaT} the percentage deviation of the single muon
flux, $\Delta I_\mu/I^0_\mu=(I_\mu-I^0_\mu)/I^0_\mu$, is shown as a function of 
the relative effective temperature variation, 
$\Delta T_{eff}/T^0_{eff}=(T_{eff}-T^0_{eff})/T^0_{eff}$. 
According to the model described in Appendix \ref{sec:Appendix}, a
proportionality relation is expected: 
\begin{equation} 
\frac{\Delta I_\mu (t)}{I_\mu^0} = \alpha_T ~\frac{\Delta T_{eff} (t)}{T_{eff}^0} 
\label{eqalphat}
\end{equation} 
The effective temperature coefficient $\alpha_T$ depends 
on the energy threshold for atmospheric muons to reach the detector 
and on the ratio of kaon/pion production in cosmic rays interactions with
the atmosphere.
The linear fit performed on the plot of Fig.~\ref{fig:alphaT}
gives $\alpha_T=0.95\pm0.04$, 
consistent with expectations and with results 
from other LNGS experiments (\cite{bib:macro1}-\cite{bib:gerda}). 
In the fit, a constant term, $a_0$, has been added 
to verify the
consistency of our data with Eq.~\ref{eqalphat}. 
Its measured value  $a_0 = -0.08 \pm 0.05$ is consistent with zero, as expected.
The correlation coefficient is $R=0.50$.  
It is worth noticing that $R$ is proportional to the cross correlation function
evaluated for $\tau=0$, and can be obtained from Eq.~\ref{eqtimecor}
replacing $I^1_\mu$ and $T^1_{eff}$, the modulation amplitudes, by the
respective standard deviations. 

Two sources of systematic errors have been investigated: the energy
threshold on cosmic ray muons collected by the 
detector, affecting
the calculation of the effective atmospheric temperatures, and the
muon rate data-set rescaling. 
Using $\langle E_{thr} \cos \theta \rangle \sim 1.8$~TeV, as done by other LNGS experiments, 
the weights 
have been re-evaluated 
resulting 
in an effective temperature which is, on average, 
0.2~K higher, with no appreciable
effect on $\alpha_T$ measurement and also on the other analyses reported here.

The evaluation of $\alpha_T$ has been performed also
applying scaling factors, different year by year, to the muon flux 
measurements, 
independently 
with and without the RPC system 
as done in Sec.~\ref{sec:Cray}; possible effects
due to small changes in detector settings and data acquisition 
would be taken into account.
The obtained effective temperature coefficient 
$\alpha_T = (0.93 \pm 0.04)$ 
is compatible with the previously
quoted value within statistical uncertainties. 
The systematic error on $\alpha_T$ has been therefore neglected with 
respect to the statistical one. 

\begin{figure}
\centering
\includegraphics[width=0.8\linewidth]{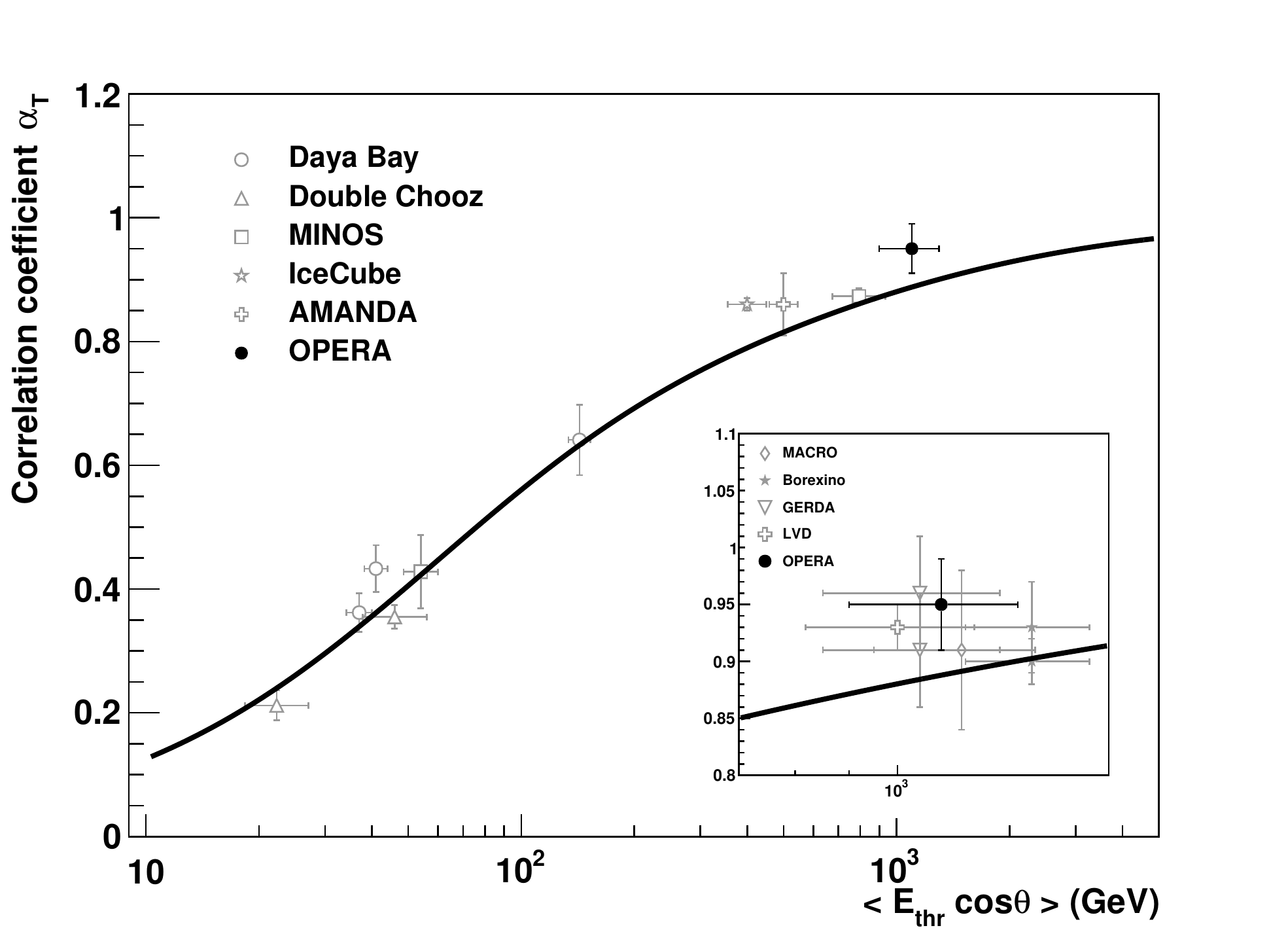}
\caption[]{ Experimental values of the correlation coefficient $\alpha_T$ as a function of $\langle E_{thr} \cos \theta \rangle$. The result of this analysis is displayed in black, measurements 
reported by other experiments in grey (AMANDA~\cite{bib:amanda}, IceCube~\cite{bib:icecube}, MINOS~\cite{bib:minosf, bib:minosn}, Double Chooz~\cite{bib:db} and Daya Bay~\cite{bib:dayabay}). 
The seven measurements performed by five LNGS 
experiments (MACRO~\cite{bib:macro1bis}, Borexino~\cite{bib:borex, bib:borex2}, GERDA~\cite{bib:gerda}, LVD~\cite{bib:lvd} and OPERA) are shown in the insert plot. 
The LNGS energy threshold - assumed to be $(1.1 \pm 0.2)$~TeV, as evaluated in this paper - is artificially displaced for each data point 
for the sake of visualisation. 
The continuous black line represents the model 
accounting for pions and kaons, with $r_{K/\pi} = 0.144$~\cite{bib:chargeratio2014}. } 
\label{fig:alphaTvsE}
\end{figure}

In Fig.~\ref{fig:alphaTvsE} our result is shown together with
the values measured by other experiments as a function of the energy threshold $\langle E_{thr} \cos \theta \rangle$. 
The model accounting for pion and kaon contributions to the muon flux is represented by the continuous black line, where the kaon/pion ratio 
$r_{K/\pi} = Z_{NK}/Z_{N\pi} = 0.144$ is fixed to the value inferred by the muon charge ratio analysis~\cite{bib:chargeratio2014}. 

\section{Conclusions}
In this paper we report on studies about the seasonal variation of 
the flux of single muons generated 
in cosmic rays interactions in the high atmosphere, as measured by the OPERA electronic detectors located in the LNGS underground laboratory. 
The observed dependence is 
approximated by a 
constant 
flux 
with the presence of a 
$(1.55 \pm 0.08) \%$ modulated component 
with one year period 
and maximum   
at 
$(186 \pm 3_{\textrm{stat}} \pm 5_{\textrm{sys}})$~day, corresponding to July 5. 

An effective atmospheric temperature has been defined as the average of the 
temperatures of the air column above the experimental site, weighted by the
production probability of high energy muons, detectable by our underground
detector.
The effective atmospheric temperature 
shows a time variation similar to cosmic ray rate changes. 

A cross correlation function based study has demonstrated the presence of 
short term deviations from a pure yearly modulated model, both in the
cosmic ray flux and in the effective temperature.  
These short term deviations and the modulated components appear to be 
simultaneous in the two time series.

The effective atmospheric temperature and muon rate variations are positively
correlated ($R=~0.50$). The effective temperature
coefficient is measured to be $\alpha_T = (0.95 \pm 0.04)$, consistent
with the model of $\pi$ and $K$ production in primary cosmic ray 
interactions and with previous experimental results. Both the phase and the effective temperature coefficient were measured by several LNGS experiments over different time periods in different decades, and the results agree very well within the quoted uncertainties. 

\appendix
\section{Models for modulation of cosmic rays and effective atmospheric 
temperature}
\label{sec:Appendix}
In this Appendix the correlation between the annual modulations of the
temperature and of the cosmic ray single muon flux is discussed according to
models described in \cite{bib:Grashorn}.

Isolated muons are detected by underground experiments as a consequence of 
decays of ($\pi^\pm$ and $K^\pm$) mesons produced by interactions of cosmic ray
primaries in the atmosphere.
An increase of the temperature results in a decrease of the air density and in 
a higher mean free path of ($\pi^\pm$ and $K^\pm$) mesons, with an increase
of single muon flux, related to the temperature variation $\Delta$T, given by:
\begin{equation} 
\label{eq:muflux} 
\Delta I_\mu (t)= \int_0^\infty dX~W(X)~ \Delta T(X,t)
\end{equation} 
where the integral extends over the atmospheric depth and $W(X)$ reflects the 
altitude dependence of the mesons production and their decays into muons that 
can be observed underground.

A suitable model \cite{bib:Grashorn} for the purpose of this paper describes 
the atmosphere as an isothermal body with an effective temperature $T_{eff}$ 
defined as:
\begin{equation} 
T_{eff} (t)= \frac {\int_0^\infty dX~T(X,t)~W(X)}{\int_0^\infty dX~W(X)}
\simeq \frac {\Sigma_{n=0}^N \Delta X_n ~T(X_n,t)~W(X_n)}{\Sigma_{n=0}^N \Delta X_n~W(X_n)}
\end{equation} 
where the approximation reflects the fact that the temperature is
measured at discrete atmospheric levels $X_n$ (not equally spaced).

Eq.~\ref{eq:muflux} can be re-written as:
\begin{equation} 
\Delta I_\mu (t)= \Delta T_{eff}(t) \int_0^\infty dX~W(X) 
\end{equation} 
or 
\begin{equation} 
\label{alphadef} 
\frac{\Delta I_\mu (t)}{I_\mu^0} = \alpha_T ~\frac{\Delta T_{eff} (t)}{T_{eff}^0}
\end{equation} 
with $\alpha_T$, the effective temperature coefficient, defined as 
\begin{equation} 
\alpha_T = \frac{T_{eff}^0}{I_\mu^0} \int_0^\infty dX~W(X) 
\end{equation} 
$T_{eff}^0$ being the effective temperature value for which $I_\mu(t)=I_\mu^0$.

The weight $W(X)$ used in Eq.~\ref{eq:muflux} can be expressed as
the sum $W_\pi + W_K$, with $W_\pi$ and $W_K$ being the contributions
of pions and kaons to the muon flux:
\begin{equation} 
\label{equationW} 
W^{\pi,K}(X) \simeq 
\frac{(1-X/\Lambda_{\pi,K}^{'})^{2}~e^{-X/\Lambda_{\pi,K}}~A_{\pi,K}^{1}}
{\gamma + (\gamma+1)~B_{\pi,K}^{1}~K(X)~(\langle E_{thr} \cos \theta \rangle/\epsilon_{\pi,K})^{2}}
\end{equation} 
where: 
\begin{equation} 
K(X) = \frac {(1-X/\Lambda_{\pi,K}^{'})^{2}} 
{(1-e^{-X/\Lambda_{\pi,K}^{'}})~\Lambda_{\pi,K}^{'}/X} 
\end{equation} 
$A_{\pi,K}^1$ and $B_{\pi,K}^1$ are constants.
$A_{\pi}^1$ is fixed to 1, while the other parameters depend on the masses of mesons and
muons, as well as on the muon spectral index $\gamma$.
$A_{K}^1$ is proportional to the ratio of forward muon production in decays
of kaons and pions, $r_{K/\pi}$.
$1/\Lambda_{\pi,K}^{\prime} = 1/\Lambda_N - 1/\Lambda_{\pi,K}$, with 
$\Lambda_N$, $\Lambda_\pi$ and $\Lambda_K$ being the attenuation lengths for
primaries, pions and kaons, respectively.
$E_{thr}$ is the energy required for a muon to reach the underground
site and $\theta$ is the angle between the muon and the vertical direction.
$\epsilon_{\pi,K}$ are the meson critical energies, for which decay
and interaction have an equal probability.  

The value of $\langle E_{thr} \cos \theta \rangle$ 
was estimated using a full Monte Carlo simulation taking into account the rock map 
above the Hall C of LNGS~\cite{chargeratio}. 
For OPERA 
$E_{thr} = 1.4$~TeV and $\langle E_{thr} \cos \theta \rangle = 1.1$~TeV.   
All other parameters are site independent. The values used in the present 
analysis are taken from~\cite{gaisser, bib:Grashorn} 
and 
are reported in Table~\ref{tableval} for the 
sake of completeness. 

\begin{figure}[]
\centerline{\includegraphics[width=0.7\linewidth]{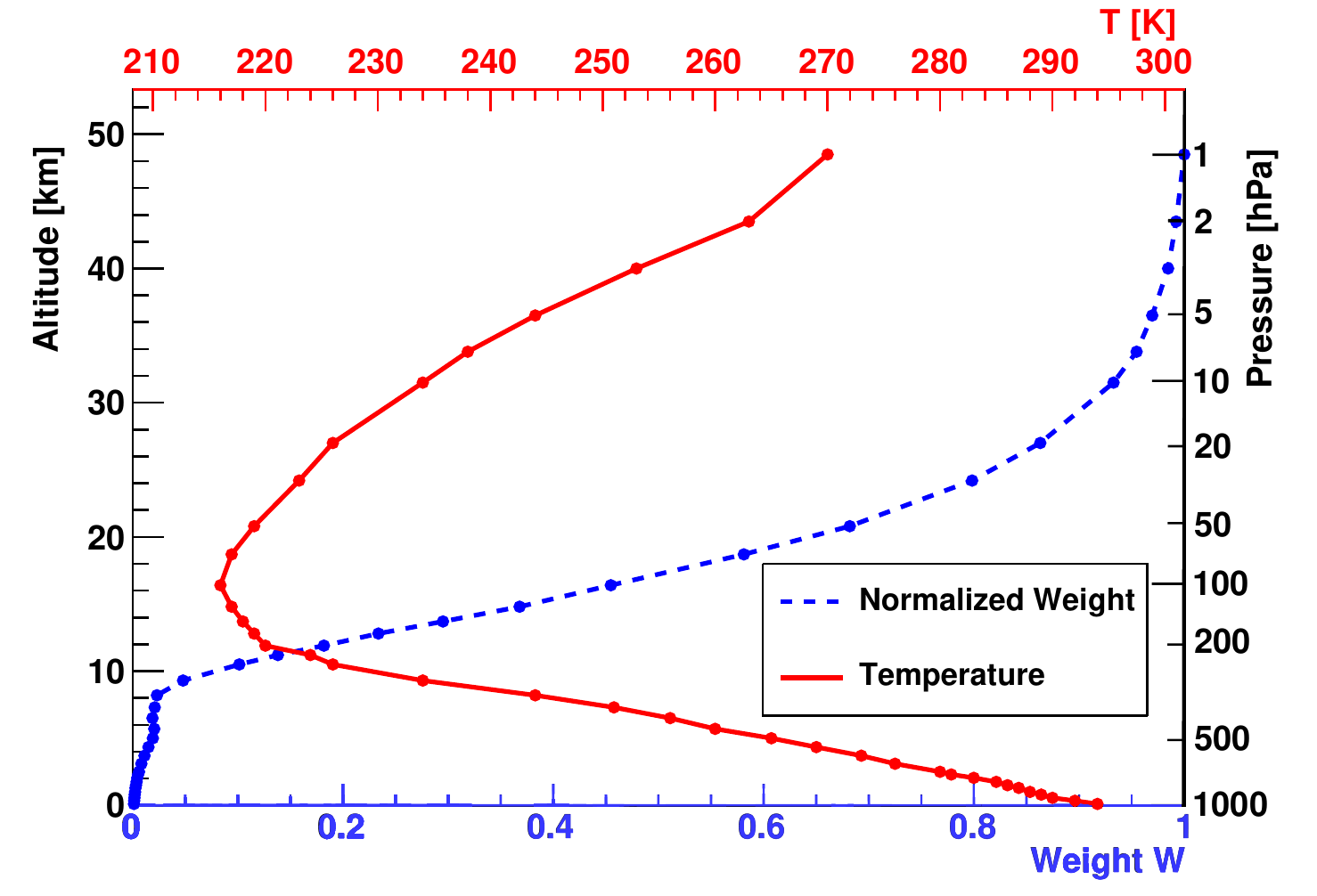}}
\caption[]{Average temperature (continuous red line) and normalized weight $W$ (dashed blue line) 
as a function of pressure levels computed for the LNGS site (levels indicated by the dots). 
In the left vertical axis, the altitude corresponding to the pressure in the right vertical axis is shown; 
the conversion 
is extracted from~\cite{bib:ECMWF} and the relationship is non-linear. } 
\label{fig:weight}
\end{figure}

\begin{table}
\centering
\caption{Input parameters used for $W(X)$ evaluation.}
\label{tableval}
\begin{tabular}{|c|c|}
\hline
Parameter & Value \\
\hline
$A_{\pi}^1$ & 1 \cite{bib:Grashorn}\\
$A_{K}^1$ & 0.38~$r_{K/\pi}$ \cite{bib:Grashorn} \\
$r_{K/\pi}$ & 0.149 \cite{gaisser}\\
$\gamma$ & 1.7 \cite{gaisser} \\
$B_{\pi}^1$ & 1.460 \cite{bib:Grashorn} \\
$B_{K}^1$ & 1.740 \cite{bib:Grashorn} \\
$\Lambda_N$ & 120 g/cm$^2$ \cite{gaisser} \\
$\Lambda_\pi$ & 160 g/cm$^2$ \cite{gaisser} \\
$\Lambda_K$ & 180 g/cm$^2$ \cite{gaisser} \\
$\epsilon_{\pi}$ & 0.115 TeV \cite{gaisser} \\
$\epsilon_{K}$ & 0.850 TeV \cite{gaisser} \\
\hline
\end{tabular}
\end{table}
In Fig.~\ref{fig:weight} the five year (2008 - 2013) average 
temperature 
in the atmosphere
for the LNGS site and the weight 
evaluated 
with $\langle E_{thr} \cos \theta \rangle = 1.1$~TeV 
as a function of the
atmospheric pressure are shown. 
The weights are normalized to the highest value ($W(\textrm{1 hPa}) = 0.0085748$). 
Higher atmosphere layers are given 
higher weights, since muons produced at low altitudes are on average
less energetic and a larger fraction of them are below the threshold
$E_{thr}$ to cross the rock overburden.

\acknowledgments
We are grateful to S. Cecchini for very useful discussions and suggestions. 
We warmly thank CERN for the successful operation of the CNGS facility and INFN for the continuous support given by hosting the experiment in its LNGS laboratory. Funding is gratefully acknowledged from national agencies and Institutions supporting us, namely: Fonds de la Recherche Scientifique-FNRS and Institut Interuniversitaire des Sciences Nucleaires for Belgium; MoSES for Croatia; CNRS and IN2P3 for France; BMBF for Germany; INFN for Italy; JSPS, MEXT, the QFPU-Global COE program of Nagoya University, and Promotion and Mutual Aid Corporation for Private Schools of Japan for Japan; SNF, the University of Bern and ETH Zurich for Switzerland; the Russian Foundation for Basic Research (Grant No. 12-02-12142 ofim), the Programs of the Presidium of the Russian Academy of Sciences (Neutrino Physics and Experimental and Theoretical Researches of Fundamental Interactions, Physics of Fundamental Interactions and Nuclear Technologies), and the Ministry of Education and Science of the Russian Federation for Russia, the Basic Science Research Program through the National Research Foundation of Korea (NRF) funded by the Ministry of Science and ICT (Grant No. NRF-2018R1A2B2007757) for Korea; and TUBITAK, the Scientific and Technological Research Council of Turkey for Turkey (Grant No. 108T324).



\begin{thebibliography}{99}




\bibitem{bib:season} 
P. H. Barret, \emph{Interpretation of cosmic-ray measurement far underground}, \emph{Rev. Mod. Phys.} {\bf 24} (1952) 133.

\bibitem{bib:macro1} 
The MACRO Collaboration, M. Ambrosio et al., \emph{Seasonal variations in the underground muon intensity as seen by MACRO}, \emph{Astropart. Phys.} {\bf 7} (1997) 109.  

\bibitem{bib:macro1bis} 
The MACRO Collaboration, M. Ambrosio et al., \emph{ Search for the sidereal and solar diurnal modulations in the total MACRO muon data set }, \emph{Phys. Rev. D} {\bf 67} (2003) 042002.   

\bibitem{bib:macro2} 
F.~Ronga, \emph{Seasonal variations of the rate of multiple-muons in the Gran Sasso underground laboratory}, EPJ Web of Conferences {\bf 136} 05004 (2017). 

\bibitem{bib:lvd} 
M.~Selvi on behalf of the LVD Collaboration, \emph{Analysis of the seasonal modulation of cosmic ray muon flux in the LVD detector during 2001-2008}, Proceedings of the $31^{st}$ International Cosmic Ray Conference, July 7-15, Lodz, Poland (2009). \\ 

C.~Vigorito on behalf the LVD collaboration, \emph{Underground flux of atmospheric muons and its variations with 25 years of data of the LVD experiment}, Proceedings of the $35^{th}$ International Cosmic Ray Conference, July 12-20, Bexco, Busan, Korea (2017). 

\bibitem{bib:borex} 
The Borexino Collaboration, G. Bellini et al., \emph{Cosmic-muon flux and annual modulation in Borexino at 3800 m water-equivalent depth}, \emph{JCAP} {\bf 5} (2012) 15. 
\bibitem{bib:borex2}
The Borexino Collaboration, M. Agostini et al., \emph{Modulations of the cosmic muon signal in ten years of Borexino data}, 
[arXiv:1808.04207]. 

\bibitem{bib:gerda} 
The GERDA Collaboration, M. Agostini et al., \emph{Flux Modulations seen by the Muon Veto of the GERDA Experiment}, \emph{Astropart. Phys.} {\bf 84} (2016) 29. 

\bibitem{bib:amanda} 
A. Bouchta for the AMANDA Collaboration, \emph{Seasonal variation of the muon flux seen by AMANDA}, Proceedings of the $26^{th}$ International Cosmic Ray Conference, 2 (1999) 108. 

\bibitem{bib:icecube} 
The IceCube Collaboration, P. Desiati et al., \emph{Seasonal variations of high energy cosmic ray muons observed by the IceCube observatory as a probe of kaon/pion ratio}, Proceedings of the $32^{nd}$ International Cosmic Ray Conference, August 11-18, Beijing, China (2011).

\bibitem{bib:minosf} 
The MINOS Collaboration, P. Adamson et al., \emph{Observation of muon intensity variations by season with the MINOS far detector}, \emph{Phys. Rev. D} {\bf 81} (2010) 012001.  
\bibitem{bib:minosn} 
The MINOS Collaboration, P. Adamson et al., \emph{Observation of muon intensity variations by season with the MINOS near detector}, \emph{Phys. Rev. D} {\bf 90} (2014) 012010. 

\bibitem{bib:db} 
The Double Chooz Collaboration, T. Abrahao et al., \emph{Cosmic-muon characterization and annual modulation measurement with Double Chooz detectors}, \emph{JCAP} {\bf 2} (2017) 17. 

\bibitem{bib:dayabay} 
The Daya Bay Collaboration, F. P. An et al., \emph{Seasonal variation of the underground cosmic muon flux observed at Daya Bay}, \emph{JCAP} {\bf 1} (2018) 1. 

\bibitem{bib:ssw} 
S.~Osprey et al., \emph{Sudden stratospheric warmings seen in MINOS deep underground muon data}, \emph{Geophys. Res. Lett.} {\bf 36} (2009) L05809.

\bibitem{bib:detector} 
The OPERA Collaboration, R. Acquafredda et al., \emph{The OPERA experiment in the CERN to Gran Sasso neutrino beam}, \emph{JINST} {\bf 4} (2009) 1.

\bibitem{bib:prl2015}
The OPERA Collaboration, N. Agafonova et al., \emph{Discovery of $\tau$ neutrino appearance in the CNGS neutrino beam with the OPERA experiment}, 
\emph{Phys. Rev. Lett.} {\bf 115} (2015) 121802. 

\bibitem{bib:prl2018}
The OPERA Collaboration, N. Agafonova et al., \emph{Final results of the OPERA experiment on $\nu_{\tau}$ appearance in the CNGS neutrino beam}, 
\emph{Phys. Rev. Lett.} {\bf 120} (2018) 211801. 

\bibitem{bib:TTref} 
T. Adam et al., \emph{The OPERA experiment Target Tracker}, 
\emph{Nucl. Instr. Meth., A} {\bf 577} (2007) 523. 

\bibitem{bib:RPC} A. Bergnoli et al., \emph{Tests of OPERA RPC Detectors}, 
\emph{IEEE Trans. Nucl. Sci.} {\bf 52} (2005) 2963. 

\bibitem{oprec} 
The OPERA Collaboration, N. Agafonova et al., \emph{Study of neutrino interactions with the electronic detectors of the OPERA experiment}, \emph{New J. Phys.} {\bf 13} (2011) 053051. 

\bibitem{bib:LS1} 
N. Lomb, \emph{Least-squares frequency analysis of unequally spaced data}, \emph{Astrophys. Space Sci.} {\bf 39} (1976) 447.

\bibitem{bib:LS2} 
J. Scargle, \emph{Studies in astronomical time series analysis. 2. Statistical aspects of spectral analysis of unevenly spaced data}, \emph{Astrophys. J.} {\bf 263} (1982) 835. 

\bibitem{bib:LSgen} 
M. Zechmeister and M. Kurster, \emph{The generalised Lomb-Scargle periodogram. A new formalism for the floating-mean and Keplerian periodograms}, \emph{Astron.Astrophys.} {\bf 496} (2009) 577. 

\bibitem{bib:ECMWF}
ECMWF, European centre for medium-range weather forecasts, http://www.ecmwf.int

\bibitem{bib:Grashorn}
E. W. Grashorn et al., \emph{The atmospheric charged kaon/pion ratio using
seasonal variation methods}, \emph{Astropart. Phys.} {\bf 33} (2010) 140. 

\bibitem{chargeratio} 
The OPERA Collaboration, N. Agafonova et al., \emph{Measurement of the atmospheric muon charge ratio with the OPERA detector}, \emph{Eur. Phys. J. C} {\bf 67} (2010) 25. 

\bibitem{bib:chargeratio2014} 
The OPERA Collaboration, N. Agafonova et al., \emph{Measurement of the TeV atmospheric muon charge ratio with the complete OPERA data set}, \emph{Eur. Phys. J. C} {\bf 74} (2014) 2933. 

\bibitem{gaisser} T. K.~Gaisser, \emph{Cosmic rays and particle physics}, Cambridge University Press (1990). 

\end{thebibliography}
\end{document}